\documentclass{emulateapj}
\usepackage{graphicx}
\usepackage{subfigure}
\newcommand{\secpoint}{\mbox{$''\mskip-7.6mu.\,$}}

\slugcomment{Received DATE; accepted DATE}

\begin{document}

\title{The Rest Frame Ultraviolet Spectra of UV-Selected Active Galactic Nuclei at $z \sim 2-3$\altaffilmark{1}}

\shorttitle{UV-SELECTED AGN COMPOSITE SPECTRA}
\shortauthors{HAINLINE ET AL.}

\author{\sc Kevin N. Hainline, Alice E. Shapley\altaffilmark{2}}
\affil{Department of Astronomy, University of California,
Los Angeles, 430 Portola Plaza, Los Angeles, CA 90024}

\author{\sc Jenny E. Greene}
\affil{Department of Astronomy, University of Texas, Austin, TX 78712}

\author{\sc Charles C. Steidel}
\affil{California Institute of Technology, MS 105-24, Pasadena, CA 91125}

\author{}

\altaffiltext{1}{Based, in part, on data obtained at the W.M. Keck
Observatory, which is operated as a scientific partnership among the
California Institute of Technology, the University of California, and
NASA, and was made possible by the generous financial support of the W.M.
Keck Foundation.}
\altaffiltext{2}{David and Lucile Packard Fellow}

\begin{abstract}

We present new results for a sample of 33 narrow-lined UV-selected active galactic nuclei (AGNs), 
identified in the course of a spectroscopic survey for star-forming galaxies at $z \sim 2-3$. 
The rest-frame UV composite spectrum for our AGN sample shows several emission lines characteristic 
of AGNs, as well as interstellar absorption features detected in star-forming LBGs. We report a detection
of \ion{N}{4}] $\lambda$1486, which has been observed in high-redshift radio galaxies, as well as in 
rare optically-selected quasars. The UV continuum slope of the composite spectrum is significantly 
redder than that of a sample of non-AGN UV-selected star forming galaxies. Blueshifted \ion{Si}{4} 
absorption provides evidence for outflowing highly-ionized gas in these objects at speeds of 
$\sim 10^{3}$ km s$^{-1}$, quantitatively different from what is seen in the outflows of non-AGN LBGs. 
Grouping the individual AGN by parameters such as Ly$\alpha$ equivalent width, redshift, and UV continuum 
magnitude allows for an analysis of the major spectroscopic trends within the sample. Stronger 
Ly$\alpha$ emission is coupled with weaker low-ionization absorption, which is similar to what is 
seen in the non-AGN LBGs, and highlights the role that cool interstellar gas plays in the escape of 
Ly$\alpha$ photons. However, the AGN composite does not show the same trends between Ly$\alpha$ 
strength and extinction seen in the non-AGN LBGs. These results represent the first 
such comparison at high-redshift between star-forming galaxies and similar galaxies that host AGN
activity. 

\end{abstract}


\section{Introduction}
\label{sec:intro}

In order to explain observations of massive galaxy evolution in the universe, current models of 
galaxy formation require a form of energetic feedback that is thought to result from the 
effects of a central active galactic nucleus (AGN) \citep{croton2006,somerville2008,dimatteo2008}. 
Energy and momentum input from the AGN into the galaxy's interstellar medium (ISM) can serve to heat
or remove gas so that it is no longer available for star formation. ``AGN feedback'' has been presented
as one of the major factors giving rise to the red sequence in massive galaxies 
\citep{silk1998,kauffmann2003b}.  Furthermore, every galaxy bulge appears to contain a supermassive 
black hole \citep{kormendy1995}, whose mass is correlated with bulge properties such as 
stellar velocity dispersion \citep{merritt2001,gultekin2009}. These correlations offer evidence 
of coupling between the formation of the black hole and bulge, which may result from the effects of 
AGN feedback \citep{silk1998,murray2005,hopkins2008}. Outflows have been observed in strongly star-forming 
galaxies over a range of redshifts \citep{franx1997,shapley2003,martin2005,steidel2010,rupke2005}, 
and post-starburst galaxies at $z \sim 0.6$ \citep{tremonti2007}. For 
AGNs, outflows have been observed in local Seyfert galaxies \citep{crenshaw2003,krug2010}, and at 
higher redshifts in broad absorption line quasars \citep{korista2008,ganguly2008}, radio galaxies 
\citep{nesvadba2006,nesvadba2008}, and ULIRGs \citep{alexander2010}. However, at early times, 
outflows have thus far not been fully examined in a sample of active galaxies that can be 
quantitatively compared to a non-active sample with similar host galaxy properties. This paper 
examines the outflow properties of such a sample at $2 \leq z \leq 3$, when both star-formation 
density and black hole (BH) accretion were at their peak \citep{madau1996, ueda2003, richards2006, 
reddy2008, silverman2008}.

The rest-frame UV portion of a galaxy spectrum is ideally suited for the study of the ISM. 
In star-forming galaxies, this spectral region contains emission or absorption from \ion{H}{1}
Lyman $\alpha$, as well as low- and high-ionization metal absorption lines that 
have been used to infer the presence of outflows \citep{pettini2000,pettini2001,pettini2002,shapley2003}.  
At $z \sim 2 - 3$, the rest-frame UV part of the spectrum is shifted into the observed optical, and
is accessible using ground-based facilities. At these redshifts, individual galaxy spectra have 
low continuum signal-to-noise (S/N), which makes robust absorption line measurements challenging. 
With a large enough sample, however, a higher S/N composite spectrum can be created, allowing 
measurements of the global properties and spectral trends within the sample. \citet{shapley2003} and 
\citet{steidel2010} have used such composite spectra to explore the outflow properties of UV-selected 
star-forming galaxies at $z \sim 2 - 3$.

Because the black hole accretion disk and broad-line region are obscured from view, the light from 
a narrow-lined AGN is not dominated by emission from the central source, but rather that of the 
host galaxy. The ability to study the host
galaxy allows for a comparison between the galaxy-scale properties of a sample of narrow-lined AGNs
and those of a similar non-AGN sample. In order to undertake such a study, we augment 
the sample of narrow-lined UV-selected AGNs at $z \sim 3$ presented in 
\citet{steidel2002}, extending it to include objects at $z \sim 2$. The original sample enabled, 
for the first time, an estimation of the fraction of star forming galaxies within the Lyman Break
Galaxy (LBG) survey that showed evidence for AGN activity on the basis of their rest frame UV 
spectroscopic properties. With the expanded sample of AGNs, we construct composite spectra that reveal the properties of outflowing gas
in these objects. The galaxies that harbor these narrow-lined AGN were selected on the basis of their broadband 
rest-frame UV colors, and should have host galaxies similar to those of the non-AGN LBGs. 
As this AGN sample appears to be hosted by galaxies drawn from the same parent population as the non-AGN LBG sample 
\citep{steidel2002,adelberger2005d}, we can conduct a controlled experiment to understand how the AGN impacts the 
gas properties of the host galaxy.

The sample of UV-selected AGNs is presented in \S \ref{sec:sample}, while in \S \ref{sec:generating} we describe
the creation of the AGN composite spectrum. This spectrum and its basic properties are shown in \S 
\ref{sec:features}, including the detection of blueshifted high-ionization absorption features. 
In \S \ref{sec:trends}, we examine spectral trends within the AGN sample that are highlighted by 
separating objects according to Ly$\alpha$ equivalent width (EW), UV magnitude, and redshift. We conclude 
in \S \ref{sec:discussion} with a discussion placing the results from the composite spectrum analysis into the context of 
our understanding of AGNs. Throughout our analysis, we assume 
$\Omega_\mathrm{M} = 0.27$, $\Omega_\Lambda = 0.73$, and $H_0 = 71$ km s$^{-1}$ Mpc$^{-1}$.

\section{The UV-Selected AGN Sample and Observations}
\label{sec:sample}

	\begin{figure}
	\epsscale{1.2} 
	\plotone{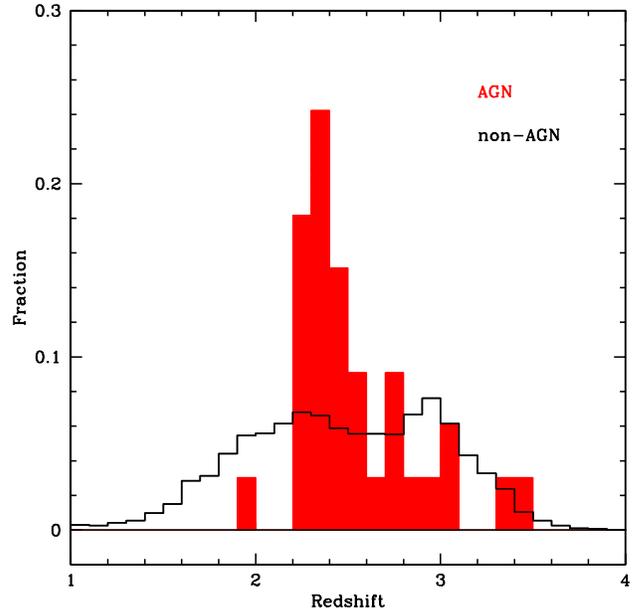} 
	\caption{The normalized redshift distribution for our sample of UV-selected narrow-lined 
	AGNs, compared to the redshifts of the $z \sim 2-3$ UV-selected non-AGN sample.
	\label{fig:redshift_dist}} 
	\epsscale{1.2}
         \end{figure}

\begin{deluxetable*}{llllcccccc}
\tabletypesize{\scriptsize}
\tablecaption{UV-Selected Narrow-Lined AGN Sample\label{tab:sample}}
\tablewidth{0pt}
\tablehead{
\colhead{FIELD} & \colhead{OBJECT} & \colhead{RA} & \colhead{DEC} & \colhead{$\cal R$} &
\colhead{$z_{Ly\alpha}$\tablenotemark{a}} & \multicolumn{4}{c}{Emission Line Detections} \\
\colhead{} & \colhead{} & \colhead{(J2000)} & \colhead{(J2000)} & \colhead{(mag)} &
\colhead{} & \multicolumn{4}{c}{}
}
\startdata

$\mathrm{Q0000}$ & $\mathrm{C7}$ & 00:03:28.85 & -26:03:53.3 & 24.21 & 3.431 & NV & CIV & HeII & CIII] \\
$\mathrm{Q0000}$ & $\mathrm{C14}$ & 00:03:30.39 & -26:01:20.7 & 24.47 & 3.057 & NV & CIV & HeII & n.c.\tablenotemark{b} \\
$\mathrm{CDFb}$ & $\mathrm{D3}$ & 00:53:43.02 & \phm{-}12:22:02.5 & 24.75 & 2.777 & NV & CIV & HeII & CIII] \\
$\mathrm{Q0100}$ & $\mathrm{BX172}$ & 01:03:08.46 & \phm{-}13:16:41.7 & 23.50 & 2.312 & NV & CIV & HeII & n.c. \\
$\mathrm{Q0142}$ & $\mathrm{BX186}$ & 01:45:17.47 & -09:45:08.0 & 25.18 & 2.361 & ...\tablenotemark{c} & CIV & ... & n.c. \\
$\mathrm{Q0142}$ & $\mathrm{BX195}$ & 01:45:17.68 & -09:44:54.2 & 23.56 & 2.382 & NV & CIV & HeII & n.c. \\
$\mathrm{Q0142}$ & $\mathrm{BX256}$ & 01:45:15.74 & -09:42:12.5 & 23.91 & 2.321 & NV & n.c. & n.c. & n.c. \\
$\mathrm{Q0201}$ & $\mathrm{oC12}$ & 02:03:56.16 & \phm{-}11:36:30.1 & 24.83 & 2.357 & NV & CIV & HeII & ... \\
$\mathrm{Q0256}$ & $\mathrm{md37}$ & 02:59:02.21 & \phm{-}00:12:03.4 & 24.06 & 2.803 & NV & CIV & ... & CIII] \\
$\mathrm{Q0933}$ & $\mathrm{MD38}$ & 09:33:48.60 & \phm{-}28:44:32.3 & 22.61 & 2.763 & ... & CIV & HeII & n.c. \\
$\mathrm{Q1217}$ & $\mathrm{BX46}$ & 12:19:19.94 & \phm{-}49:40:22.7 & 23.85 & 1.980 & NV & ... & ... & n.c. \\
$\mathrm{HDF}$ & $\mathrm{MMD12}$ & 12:37:19.80 & \phm{-}62:09:56.0 & 24.36 & 2.648 & NV & CIV & ... & CIII] \\
$\mathrm{HDF}$ & $\mathrm{BMZ1156}$ & 12:37:04.34 & \phm{-}62:14:46.3 & 24.62 & 2.211 & NV & CIV & HeII & CIII] \\
$\mathrm{HDF}$ & $\mathrm{BMZ1384}$ & 12:37:23.15 & \phm{-}62:15:38.0 & 23.98 & 2.243 & NV & CIV & HeII & CIII] \\
$\mathrm{HDF}$ & $\mathrm{BX160}$ & 12:37:20.07 & \phm{-}62:12:22.7 & 24.02 & 2.461 & NV & CIV & ... & ... \\
$\mathrm{Westphal}$ & $\mathrm{MM47}$ & 14:17:57.39 & \phm{-}52:31:04.5 & 24.30 & 3.027 & ... & CIV & HeII & n.c. \\
$\mathrm{Westphal}$ & $\mathrm{MMD58}$ & 14:17:18.29 & \phm{-}52:28:53.9 & 25.36 & 2.596 & ... & n.c.\tablenotemark{d} & HeII & ... \\
$\mathrm{Q1422}$ & $\mathrm{d14}$ & 14:24:47.44 & \phm{-}22:48:04.3 & 24.30 & 2.245 & NV & CIV & ... & CIII] \\
$\mathrm{Q1422}$ & $\mathrm{c73}$ & 14:24:46.41 & \phm{-}22:55:45.5 & 24.88 & 3.382 & NV & n.c. & n.c. & n.c. \\
$\mathrm{Q1422}$ & $\mathrm{md109}$ & 14:24:42.58 & \phm{-}22:54:46.6 & 23.69 & 2.229 & NV & CIV & HeII & CIII] \\
$\mathrm{Q1623}$ & $\mathrm{BX151}$ & 16:25:29.61 & \phm{-}26:53:45.0 & 24.60 & 2.441 & NV & CIV & HeII & n.c. \\
$\mathrm{Q1623}$ & $\mathrm{BX454}$ & 16:25:51.42 & \phm{-}26:43:46.3 & 23.89 & 2.422 & NV & CIV & ... & ... \\
$\mathrm{Q1623}$ & $\mathrm{BX663}$ & 16:26:04.58 & \phm{-}26:48:00.2 & 24.14 & 2.435 & NV & CIV & HeII & ... \\
$\mathrm{Q1623}$ & $\mathrm{BX747}$ & 16:26:13.46 & \phm{-}26:45:53.2 & 22.55 & 2.441 & NV & CIV & HeII & n.c. \\
$\mathrm{Q1623}$ & $\mathrm{BX827}$ & 16:26:19.31 & \phm{-}26:45:15.3 & 25.15 & 2.506 & ... & CIV & ... & n.c. \\
$\mathrm{Q1700}$ & $\mathrm{MD157}$ & 17:00:52.19 & \phm{-}64:15:29.3 & 24.35 & 2.295 & NV & CIV & HeII & CIII] \\
$\mathrm{Q1700}$ & $\mathrm{MD174}$ & 17:00:54.54 & \phm{-}64:16:24.8 & 24.56 & 2.347 & NV & ... & ... & n.c. \\
$\mathrm{Q2233}$ & $\mathrm{D3}$ & 22:36:16.12 & \phm{-}13:55:19.2 & 23.93 & 2.795 & NV & CIV & HeII & CIII] \\
$\mathrm{Q2233}$ & $\mathrm{MD21}$ & 22:36:35.83 & \phm{-}13:55:42.0 & 24.80 & 2.549 & NV & CIV & HeII & n.c. \\
$\mathrm{DSF2237a}$ & $\mathrm{D11}$ & 22:40:02.99 & \phm{-}11:52:13.9 & 25.19 & 2.959 & ... & CIV & HeII & CIII] \\
$\mathrm{DSF2237b}$ & $\mathrm{MD53}$ & 22:39:28.67 & \phm{-}11:52:09.5 & 24.07 & 2.292 & ... & CIV & HeII & CIII] \\
$\mathrm{Q2343}$ & $\mathrm{BX333}$ & 23:46:21.51 & \phm{-}12:47:03.2 & 24.12 & 2.397 & NV & CIV & HeII & CIII] \\
$\mathrm{Q2346}$ & $\mathrm{BX445}$ & 23:48:13.20 & \phm{-}00:25:15.8 & 23.66 & 2.330 & NV & CIV & HeII & CIII] \\

\enddata

\tablenotetext{a}{Emission-line redshift, as measured from the Ly$\alpha$ feature.}
\tablenotetext{b}{The feature of interest was not covered by the LRIS spectral range, indicated as ``n.c."}
\tablenotetext{c}{The feature of interest was covered by the LRIS spectral range but not significantly detected.}
\tablenotetext{d}{A sky line coincides with the wavelength of CIV.}
\end{deluxetable*}

The narrow-lined AGNs used to construct the composite spectra were discovered in the course
of a survey of $z \sim 2 - 3$ UV-selected galaxies. This survey spans 0.89 degrees$^2$ across 29 fields and 
is described in detail in \citet{steidel2003,steidel2004} and \citet{reddy2008}. The
method for selecting $z \sim 3$ LBGs is based on the fact that the intervening intergalactic
medium absorbs most of the photons with wavelengths shortward of the Lyman break at 912 \AA. At $z \sim 3$,
the Lyman break is shifted to optical wavelengths, and galaxies at this redshift are selected
by their position in a $U-G$ vs. $G-{\cal R}$ color-color diagram. As described in \citet{steidel2004} and \citet{adelberger2004a},
color criteria have also been developed to identify galaxies with similar intrinsic rest-frame UV colors to 
the $z \sim 3$ LBGs, but at $z \sim 2$. At this redshift, the observed $U G {\cal R}$ colors reflect a flat part of 
the spectrum redward of the Lyman break for star-forming galaxies. The survey of $z \sim 2 - 3$ UV-selected
galaxies is split into various subsamples based on redshift, including 
the ``BM'' ($1.5 \leq z \leq 2.0$), ``BX'' ($2.0 \leq z \leq 2.5$), and the ``C'', ``D'',  and ``MD'' galaxies 
($2.7 \leq z \leq 3.3$). From the photometric sample, objects were selected for spectroscopy without 
regard for their x-ray or morphological properties. Therefore, the fraction of candidates
targeted for spectroscopy should be similar for both AGNs and star-forming galaxies in the UV-selected
photometric sample.

The parent UV-selected spectroscopic sample consists of 3059 galaxies, 48 broad-lined AGN, and 
33 narrow-lined AGN (which comprise 1\% of the sample). Objects were identified as narrow-lined AGN if they showed 
strong Ly$\alpha$ emission accompanied by significant emission in either \ion{N}{5} $\lambda$1240 or 
\ion{C}{4} $\lambda$1549, where the FWHM for any of the emission features was less than 2000 km s$^{-1}$. In some objects, 
\ion{Si}{4} $\lambda \lambda$1393,1402 or \ion{He}{2} $\lambda$1640 was additionally used to indicate 
the presence of an AGN. In practice, the requirement of strong Ly$\alpha$ emission does not select against 
finding AGNs with strong nebular emission lines but weak Ly$\alpha$ emission. In the sample of AGNs 
presented here, Ly$\alpha$ is is on average $\sim 4-5$ times stronger than \ion{C}{4}, which is typically 
the next strongest feature. Furthermore,
no objects were found in the parent sample of UV-selected galaxies with high-ionization emission
lines while lacking Ly$\alpha$ emission. The AGN sample presented here serves as a follow-up to the 
one described in \citet{steidel2002}, now including BM/BX and additional MD objects, doubling the 
number of discovered AGN. The normalized redshift distributions for the AGN and non-AGN samples are shown in Figure 
\ref{fig:redshift_dist}. The average redshift of the 33 AGNs is $\langle z \rangle = 2.55 \pm 0.31$, 
with an average ${\cal R}$ magnitude of $\langle {\cal R} \rangle = 24.2 \pm 0.7$ (a range in 
${\cal R}$ of $22.55 - 25.72$). For the sample of non-AGNs, reflecting the combined C, D, M, MD, BX, 
and BM selection windows, the average redshift is $\langle z \rangle = 2.49 \pm 0.59$, with an 
average ${\cal R}$ magnitude of $\langle {\cal R} \rangle = 24.4 \pm 0.6$ (a range in ${\cal {\cal R}}$ 
of $21.66 - 25.97$\footnote{The standard ${\cal R} = 25.5$ limit was relaxed in the Q1422 field.}). 
Within the AGN sample, there are 2 BM objects, 13 BX, 9 MD, 4 C, 4 D, and 1 M. These represent 
$\sim 1 - 2 \%$ of each photometric class within the parent spectroscopic sample. The full sample is
listed in Table \ref{tab:sample}. 

When considering the demographics and space density of our sample of AGNs, it is important to consider 
how their spectroscopic and photometric properties affect their selection, as done for the full sample of non-AGN LBGs in 
\citet{reddy2008}. To estimate completeness for our sample, we follow the methodology of \citet{steidel2002}, 
updated for the addition of the $z \sim 2$ sample. Redshifts were measured for the sample of LBGs by virtue of 
emission and absorption features detected in the spectra. The AGN were selected based on the presence of  
emission lines in addition to Ly$\alpha$. In order to estimate an upper limit on the number of 
AGN that would be unrecognized because of low S/N spectra, we examine the strength of the strongest 
emission line used to infer the existence of an AGN, \ion{C}{4} $\lambda$1549. As discussed above, 
\ion{C}{4} is, on average, $\sim 20\%$
the strength of Ly$\alpha$ for the set of 33 AGN spectra. As reported in \citet{steidel2002}, the ratio
of \ion{C}{4} to Ly$\alpha$ emission in the $z \sim 3$ non-AGN LBG sample is $\leq$1\% for the quartile of LBGs 
having the largest Ly$\alpha$ EW. For the quartile of $z \sim 2$ non-AGN with the strongest Ly$\alpha$
emission, we measure the ratio of \ion{C}{4} to Ly$\alpha$ flux to be 1\% (Dawn Erb, private communication).
Given the observed range of Ly$\alpha$ EWs in our sample, we conclude that narrow-lined AGNs would only be found
in this strong-Ly$\alpha$ quartile. If we assume that all of the \ion{C}{4} emission in the strong-Ly$\alpha$ composite 
spectrum is originating from a small sample of undetected AGNs with \ion{C}{4}/Ly$\alpha$ ratios of 20\%, 
then the observed \ion{C}{4}/Ly$\alpha$ ratio of 1\% in the strong Ly$\alpha$ composite implies that such 
objects can only make up 5\% of the sample. If we then assume that there are no additional unrecognized AGNs 
in the other three quartiles with weaker Ly$\alpha$ emission, the fractional contribution of unrecognized AGNs to the full 
$z \sim 2$ non-AGN sample is $\sim 1\%$. Therefore, our observed frequency of $\sim 1\%$ of AGN 
is likely representative of both the spectroscopic and photometric samples of UV-selected $z \sim 2 - 3$ galaxies. 

The AGNs presented here are first identified as high-redshift galaxies based on their location 
in $U-G$ vs. $G-{\cal R}$ color-color space. Given that the average rest-frame Ly$\alpha$ emission 
EW for the AGN sample is $\langle W_{Ly\alpha} \rangle = 80$\AA, we must consider the effect of Ly$\alpha$ 
emission on the selection of these objects as a function of redshift. Within certain redshift 
ranges, Ly$\alpha$ can contribute flux to either the $U$ or $G$ bands and alter the position of an 
AGN in color-color space such that it scatters into or out of the UV-selected sample. For 19 out of 
33 AGNs in our sample at $2.17 \leq z \leq 2.48$ the Ly$\alpha$ feature does not fall 
within either the $U$ or $G$ band, and therefore Ly$\alpha$ emission contributions are irrelevant. 
One of these objects, HDF-BMZ1156, has anomalously strong \ion{C}{4} emission, which has a 
significant impact on the observed broadband rest-frame UV colors. For the remaining AGNs 
(1 object at $z\leq 2.17$, and 11 at $2.48 \leq z \leq 3.40$), we investigate how Ly$\alpha$ emission 
affects their inclusion in the UV-selected sample. Using the measured Ly$\alpha$ EW for each object, 
we correct the $U$ or $G$ band magnitude for Ly$\alpha$ flux as needed in order to characterize the 
broadband colors in the absence of line emission. This analysis reveals that only 4 out of 12 
objects were actually scattered into the selection windows due to the presence of Ly$\alpha$ 
emission. HDF-BMZ1156 was scattered into the selection window due to its \ion{C}{4} emission. 
In summary, only 5 out of our 33 AGNs have broadband colors that fail to satisfy the $UG{\cal R}$ 
selection criteria if not for the presence of strong line emission in the $U$ or $G$ filter. On the
other hand, the presence of strong Ly$\alpha$ emission may also cause objects to scatter out of the 
high-redshift galaxy color selection windows. For example, the average Ly$\alpha$ EW of the AGN sample
represents a shift of $\Delta G = 0.26$ magnitudes for an object at $z = 2.7$ (and a comparable 
shift in the $U$ band magnitude for an object at $z = 2.0$). Therefore, objects at $z>2.48$ with 
Ly$\alpha$ emission EWs of this strength will not be identified as high-redshift star-forming galaxies 
if their line-free $G-{\cal R}$ color is within $\Delta G$ of the blue $G-{\cal R}$ edge of the 
color selection box. For a robust comparison of the rest-frame UV continuum and emission-line 
properties of UV-selected AGNs and non-AGNs, the issues of completeness in color-color space as a
result of strong line emission are critical. ÊBecause emission lines do not significantly affect 
the $U$ and $G$ magnitudes of UV-selected AGNs at $2.17 \leq z \leq 2.48$, these objects serve 
as a powerful control sample when we compare the rest-frame UV properties of AGNs and non-AGNs in the
following sections.

The AGN spectra were obtained with the LRIS spectrograph at the W.M. Keck
Observatory \citep{oke1995}, using multi-object slitmasks with either 1\secpoint4 or 1\secpoint2 slits. 
The majority of the spectra in our sample were obtained with either a 300 line mm$^{-1}$ grating 
blazed at 5000 \AA\, or, following the installation of the blue channel of the LRIS instrument, a
400 groove mm$^{-1}$ grism blazed at 3400 \AA. Additionally, a small number of spectra were obtained 
with the 600 groove mm$^{-1}$ grism blazed at 4000 \AA. We refer readers to 
\citet{steidel2003,steidel2004} for a full discussion of the reduction procedures for these data. 

\section{Generating the Composite Spectrum}
\label{sec:generating}

	\begin{figure*}
	\epsscale{1.18} 
	\plotone{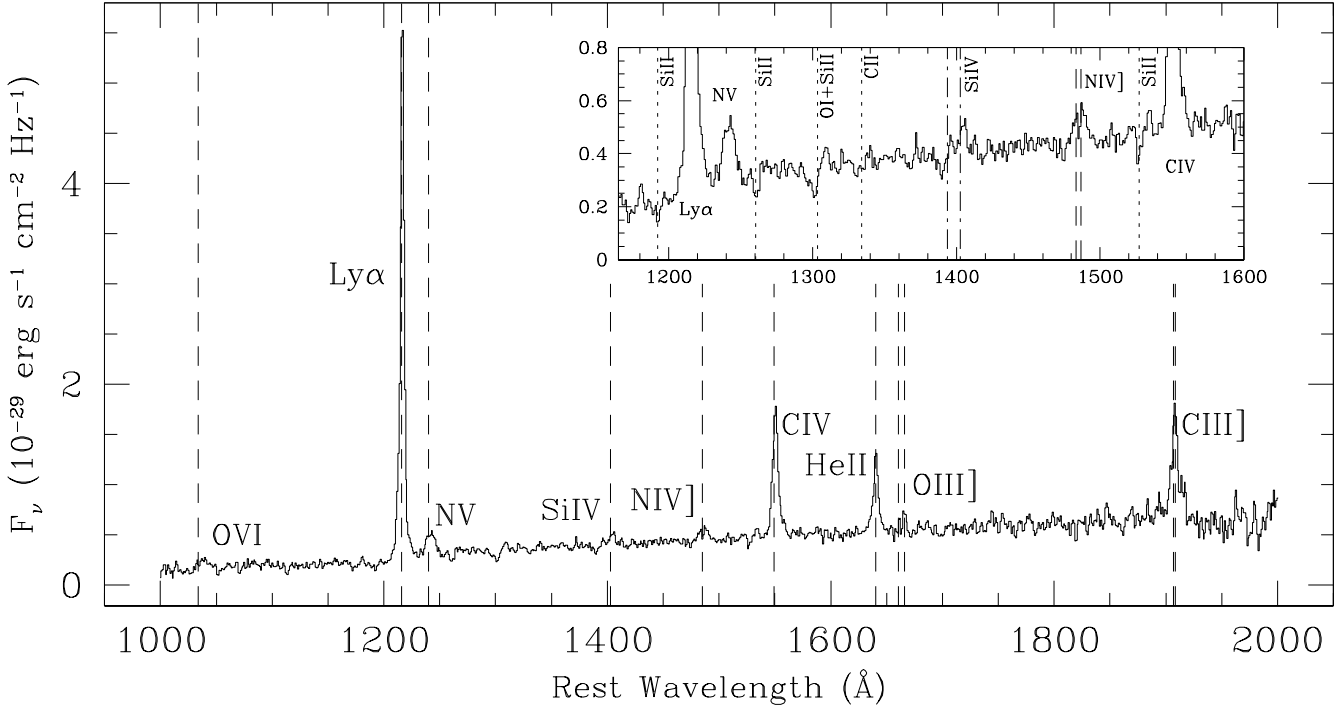} 
	\caption{The composite rest-frame UV spectrum for 33 narrow-lined AGN at $z \sim 2-3$. 
	These objects were selected by virtue of their emission lines, such as \ion{N}{5}$\lambda$1240, 
	\ion{C}{4} $\lambda$1549, and \ion{He}{2}$\lambda$1640. 
	The dashed lines indicate the locations of emission lines, while the
	inset highlights some of the more prominent absorption features. Dotted lines are used to mark 
	low-ionization absorption features, and dot-dashed lines indicate high-ionization 
	absorption features. 
	\label{fig:comp_spec}} 
	\epsscale{1.18}
         \end{figure*}

The current sample of narrow-lined UV-selected AGNs is more than twice as large as the one presented in
\citet{steidel2002}. Therefore, the resulting composite spectrum of these objects enables the 
identification of weak emission and absorption lines not visible in the average of the previous smaller sample.
To create the composite spectrum, we took the individual extracted, one-dimensional, 
flux-calibrated AGN spectra from our sample, and shifted them to the rest frame as
described below. The spectra were scaled to a common median in the wavelength range of 
$1250 - 1380$ \AA\ and then averaged. To exclude positive and negative sky subtraction 
residuals, we rejected the four highest and lowest outliers at each 
wavelength of the composite spectrum, which corresponded to $\sim$20\% of the data. 
Composite spectra constructed from the median of the flux values at each wavelength yielded results 
indistinguishable from those based on composite spectra constructed from the average of the flux values.

In order to construct the composite spectrum, each individual spectrum was shifted to the 
rest frame, which requires an accurate measurement of the galaxy's systemic redshift. The
establishment of an accurate systemic redshift is also necessary for estimating the absolute
values of kinematic offsets for different interstellar components in the composite spectrum.
The low S/N of the AGN spectra only allows for a measurement of the strongest rest-frame
UV features, such as the emission lines \ion{H}{1} Ly$\alpha$, \ion{N}{5} $\lambda$1240, 
\ion{C}{4} $\lambda$1549, \ion{He}{2} $\lambda$1640, and \ion{C}{3}] $\lambda 
\lambda$1907, 1909. Absorption features were not detected in enough spectra to be used to calculate 
redshifts. The spectra for all of our 33 objects contain a Ly$\alpha$ emission feature. However,
it is well known that Ly$\alpha$ is not a reliable indicator of the stellar redshift of a 
galaxy due to resonant scattering of the line \citep{pettini2001,shapley2003,adelberger2003,steidel2010}, 
therefore we consider the emission lines associated with AGN activity. As indicated in Table 
\ref{tab:sample}, \ion{N}{5} $\lambda$1240, \ion{C}{4} $\lambda$1549, \ion{He}{2} $\lambda$1640, 
and \ion{C}{3}] $\lambda \lambda$1907, 1909 were detected in many individual spectra. However, 
many of these detections do not have the signal-to-noise necessary to calculate an accurate redshift.
Furthermore, \ion{C}{3}] $\lambda \lambda$1907, 1909 is a density-sensitive doublet whose blended
centroid varies as a function of density, so this feature is also unsuitable for measuring 
redshifts. In general, \ion{C}{4} and \ion{He}{2} were the two emission features that yielded the 
highest signal-to-noise detections in the individual spectra. We measure accurate centroids for 
\ion{C}{4} in 19 of our 33 objects, and \ion{He}{2} in 17 of our 33 objects. We calculated how the 
redshifts derived from \ion{C}{4} $\lambda$1549 and \ion{He}{2} $\lambda$1640 
compared to the  Ly$\alpha$ redshifts. The average (median) difference in velocity between Ly$\alpha$ and 
\ion{C}{4} $\lambda$1549, is $\Delta v = -7$ km s$^{-1}$ ($24$ km s$^{-1}$), where a negative [positive]
velocity difference indicates blueshift [redshift]. It has previously been 
shown that the velocities traced by the Ly$\alpha$ and \ion{C}{4} emission lines agree in active 
galaxies \citep{buson1990}. However, the average (median) difference in velocity between 
Ly$\alpha$ and \ion{He}{2} $\lambda$1640 is $\Delta v = 173$ km s$^{-1}$ ($169$ km s$^{-1}$). 
As \ion{He}{2} $\lambda$1640 is not a resonance line, it serves as a better tracer for 
the redshift of the stars in the galaxy than either Ly$\alpha$ or \ion{C}{4} $\lambda$1549, 
which are subject to radiative transfer effects \citep{villar2000,villar2003}. 

In the absence of stellar absorption lines, H$\alpha$ is used in the rest-frame optical 
to derive redshifts. We compared the redshifts derived from optical 
spectra for five AGN from our sample taken with the NIRSPEC instrument on the Keck II telescope \citep{mclean1998}. For 
these objects, the average (median) velocity difference between Ly$\alpha$ and  H$\alpha$ is 
$\Delta v = 132$ km s$^{-1}$ ($208$ km s$^{-1}$), while the average (median) difference between \ion{He}{2} 
$\lambda$1640 and H$\alpha$ is $\Delta v = -37$ km s$^{-1}$ ($59$ km s$^{-1}$). With this in mind, 
the average velocity difference between Ly$\alpha$ and \ion{He}{2} $\lambda$1640, for
composite spectra created with only those objects where both lines were detected in the spectrum, 
is $-196 \pm 50$ km s$^{-1}$. We adopted this Ly$\alpha$-\ion{He}{2} offset for all of the objects
in the sample, and added it to each Ly$\alpha$ redshift to estimate the systemic redshift
for each object. We note that the composite spectrum is virtually identical if individual
\ion{He}{2} redshifts are used to estimate the rest frame for the objects with robust \ion{He}{2} 
centroid measurements, as opposed to the average Ly$\alpha$-\ion{He}{2} offset. Furthermore, a 
stack of the 16 spectra without individual \ion{He}{2} centroid measurements, using the average 
Ly$\alpha$-\ion{He}{2} offset to estimate the rest frame, results in a \ion{He}{2} profile centered
at zero velocity. These consistency checks indicate the validity of adopting the average 
Ly$\alpha$-\ion{He}{2} offset for all spectra in order to estimate the rest frame.

\section{Features In The Composite Spectrum}
\label{sec:features}

The final composite spectrum is shown in Figure \ref{fig:comp_spec}, an average of the spectra
for our 33 AGNs. We identify several nebular emission lines, including \ion{N}{5} $\lambda$1240,
\ion{C}{4} $\lambda$1549, \ion{He}{2} $\lambda$1640, and \ion{C}{3}] $\lambda \lambda$1907, 1909. 
We also detect \ion{N}{4}] $\lambda$1486, which has previously been detected locally in compact 
planetary nebulae \citep{davidson1986}, at $z = 3.4$ in the Lynx arc \citep{fosbury2003}, 
in high-redshift QSOs \citep{bentz2004,glikman2007}, in a $z = 5.563$ Ly$\alpha$ emitter 
\citep{vanzella2009,raiter2010}, and in radio galaxies \citep{vernet2001,humphrey2008}. We also detect absorption features 
shown in the inset of Figure \ref{fig:comp_spec}. Many of these are observed in the LBG non-AGN 
composite spectrum from \citet{shapley2003}, including the low-ionization lines, \ion{Si}{2} $\lambda$1260, \ion{O}{1} 
$\lambda$1302+\ion{Si}{2} $\lambda$1304, \ion{Si}{2} $\lambda$1527, and high-ionization feature,
\ion{Si}{4} $\lambda \lambda$1393,1402.

We measured the rest-frame centroids (and corresponding relative velocities), EWs, and FWHM
values for each of the emission lines and absorption lines, which are found in Tables \ref{tab:abs} and
\ref{tab:emi}. Gaussian profiles were fit to the measured spectroscopic features to 
estimate line centroids. These centroids were then used to estimate velocities relative to the
systemic frame. In cases where the lines are a blend of two closely spaced emission or absorption 
features, a rest-frame blend value was calculated based on the oscillator strengths for the pair
and used to estimate the velocity. The EW was calculated by first measuring
the values of the continuum on either side of the line of interest, integrating the flux 
in the line between these points, and then dividing the flux by the average of the continuum values. 

\begin{deluxetable*}{lrcccr}
\tabletypesize{\scriptsize}
\tablecaption{Absorption Features\label{tab:abs}}
\tablewidth{0pt}
\tablehead{
  & \colhead{$\lambda_{\mathrm{lab}}^a$} & \colhead{$\lambda_{\mathrm{blend}}^a$} & \colhead{$\lambda_{\mathrm{obs}}^b$} & \colhead{$W_0^{c}$} & \colhead{$\Delta v^d$} \\
\colhead{Ion} & \colhead{(\AA)} & \colhead{(\AA)} & \colhead{(\AA)} & \colhead{(\AA)} & \colhead{(km s$^{-1}$)} 
}
\startdata

$\mathrm{SiII}\tablenotemark{e}$    & 1190.42, 1193.28 & 1192.33 & 1191.84$\pm$1.32 & $-$1.70$\pm$0.68 & $-$123$\pm$333 \\
$\mathrm{SiII}\tablenotemark{e}$    & 1260.42          &    -    & 1260.45$\pm$0.94 & $-$2.51$\pm$0.55 & ~~~~~7$\pm$223 \\
$\mathrm{OI+SiII}\tablenotemark{e}$ & 1302.17, 1304.37 & 1303.27 & 1300.37$\pm$0.94 & $-$2.11$\pm$0.34 & $-$667$\pm$216 \\
$\mathrm{SiIV}\tablenotemark{f}$    & 1393.76          &    -    & 1389.83$\pm$0.83 & $-$1.18$\pm$0.44 & $-$845$\pm$178 \\
$\mathrm{SiII}\tablenotemark{e}$    & 1526.7\phm{0}           &    -    & 1526.49$\pm$1.00 & $-$0.73$\pm$0.27 & ~$-$41$\pm$197 \\

\enddata

\tablenotetext{a}{Vacuum wavelength.}
\tablenotetext{b}{Observed wavelength.}
\tablenotetext{c}{Rest-frame EW and 1$\sigma$ error. The error takes into account both sample variance
and the S/N of the composite spectrum (see \S \ref{sec:obsdata}).}
\tablenotetext{d}{Relative velocity measured in the systemic rest frame of the composite spectrum, equivalent to the rest frame of the stars.}
\tablenotetext{e}{\phm{}Low-ionization absorption feature.}
\tablenotetext{f}{High-ionization absorption feature.}

\end{deluxetable*}

\begin{deluxetable*}{lrcccr}
\tabletypesize{\scriptsize}
\tablecaption{Emission Features\label{tab:emi}}
\tablewidth{0pt}
\tablehead{
  & \colhead{$\lambda_{\mathrm{lab}}^a$} & \colhead{$\lambda_{\mathrm{blend}}^a$} & \colhead{$\lambda_{\mathrm{obs}}^b$} & \colhead{$W_0^{c}$} & \colhead{$\Delta v^d$} \\
\colhead{Ion} & \colhead{(\AA)} & \colhead{(\AA)} & \colhead{(\AA)} & \colhead{(\AA)} & \colhead{(km s$^{-1}$)} 
}
\startdata

$\mathrm{OVI}$      & 1031.912, 1037.613 & 1033.82 & 1037.94$\pm$2.20 & 6.01$\pm$2.20   & ~1196$\pm$639  \\
$\mathrm{Ly\alpha}$ & 1215.67            &    -    & 1216.47$\pm$0.04 & 66.39$\pm$11.65 & ~~197$\pm$10\phm{0}  \\
$\mathrm{NV}$       & 1238.821, 1242.804 & 1240.15 & 1242.09$\pm$0.64 & 5.60$\pm$1.00   & ~~470$\pm$154   \\
$\mathrm{SiII^{*e}}$   & 1309.276           &    -    & 1309.22$\pm$0.54 & 0.63$\pm$0.24   & $-$13$\pm$124 \\
$\mathrm{SiIV}$     & 1402.77            &    -    & 1405.02$\pm$0.93 & 1.44$\pm$0.58   & ~~481$\pm$176   \\
$\mathrm{NIV]}$     & 1483.321, 1486.496 &    \phm{$^d$}-$^d$    & 1486.59$\pm$1.17 & 2.00$\pm$0.78   &    \\
$\mathrm{CIV}$      & 1548.202, 1550.774 & 1549.06 & 1550.47$\pm$0.38 & 16.34$\pm$2.86\phm{0}  & ~~273$\pm$74\phm{0}  \\
$\mathrm{HeII}$     & 1640.405           &    -    & 1640.45$\pm$0.27 & 8.07$\pm$1.32   & ~~~~8$\pm$50\phm{0}    \\
$\mathrm{OIII]}$    & 1660.809           &    -    & 1661.63$\pm$1.28 & 0.32$\pm$1.32   & ~~148$\pm$231   \\
$\mathrm{OIII]}$    & 1666.150           &    -    & 1666.09$\pm$0.65 & 1.10$\pm$0.39   & $-$10$\pm$118 \\
$\mathrm{CIII]}$    & 1906.683, 1908.734 &    \phm{$^d$}-$^d$    & 1908.28$\pm$0.80 & 13.96$\pm$4.20\phm{0}  &   \\

\enddata

\tablenotetext{a}{Vacuum wavelength.}
\tablenotetext{b}{Observed wavelength.}
\tablenotetext{c}{Rest-frame EW and 1$\sigma$ error. The error takes into account both sample variance
and the S/N of the composite spectrum (see \S \ref{sec:obsdata}).}
\tablenotetext{d}{Relative velocity measured in the systemic rest frame of the composite spectrum, 
equivalent to the rest frame of the stars. Relative velocities were not calculated for the 
density-sensitive N~IV] and C~III] doublets, for which the blend wavelength depends on the
uncertain electron density.}
\tablenotetext{e}{The asterisk indicates that this feature is a fine-structure emission line.
}

\end{deluxetable*}

\subsection{Uncertainties}
\label{sec:obsdata}

The measurements from the composite spectrum are only meaningful if accompanied by an analysis
of the associated errors. The uncertainty on the EW values for the
composite spectrum is dependent on both the range of EW of the individual sample 
galaxies as well as the finite S/N of the composite spectrum. Almost all but the strongest 
features cannot be measured in individual spectra.

To reflect sample variance and also statistical noise, we used a bootstrap technique 
outlined in \citet{shapley2003} to calculate the uncertainties. In this method, we generated 500 
fake composite spectra constructed from the sample of spectra used in creating the real 
composite spectra. Each fake spectrum was constructed in the same way, with the same number of 
spectra as the actual composite, but with the list of input spectra formulated by selecting 
spectra at random, with replacement, from the full list of AGN spectra ($\sim36$\% of sample is 
replaced by duplicates). With these 500 fake spectra, we measured the line centroids (and 
velocities) and EWs for all of the emission and absorption features detected in the original composite spectrum. 
The standard deviation of the 500 individual measurements represents the errors on these values. 

\subsection{Emission Features}
\label{sec:emissionfeat}

The emission features for the full composite spectrum are shown in Table \ref{tab:emi}. The most
prominent feature is Ly$\alpha$, which, for the individual AGN spectra, is always observed in emission.
In \citet{shapley2003}, the Ly$\alpha$ feature was detected both in emission and absorption for a 
sample of 820 non-AGN LBGs. We measure a $\Delta v = +197$ km s$^{-1}$ emission redshift
from our composite spectrum, which is slightly lower, but similar to what is observed in the non-AGN 
LBG sample ($\Delta v = +360$ km s$^{-1}$). This redshift is due to the scattering of Ly$\alpha$ photons 
off of redshifted material in the galaxy, which allows the (now off-resonance) photons to escape. 
The measured EW is $W_{\mathrm{Ly}\alpha} = 66$ \AA, which is similar to the EW measured from the 
quartile of non-AGN objects from \citet{shapley2003} with the strongest Ly$\alpha$ EW. 

The other strong emission lines we see in the composite spectrum are \ion{N}{5} $\lambda$1240 ($W_{\mathrm{N~V}} = 5.6$ \AA), 
\ion{C}{4} $\lambda$1549 ($W_{\mathrm{C~IV}} = 16.3$ \AA), \ion{He}{2} $\lambda$1640 ($W_{\mathrm{He~II}} = 8.1$ \AA), 
and \ion{C}{3}] $\lambda \lambda$1907, 1909 ($W_{\mathrm{C~III]}} = 14.0$ \AA). The majority of non-AGN LBGs do not
show evidence for these features with such large EW values, even in those LBGs with strong Ly$\alpha$
emission \citep{shapley2003}. Both \ion{N}{5} and \ion{C}{4} have redshifts that are similar to the one measured for
Ly$\alpha$. Because we are using a composite spectrum, which has a higher S/N ratio than the individual 
spectra, we can detect weaker emission lines as well. We detect \ion{N}{4}] $\lambda$1486,
as well as \ion{O}{6} $\lambda \lambda$1032, 1038, and \ion{O}{3}] $\lambda \lambda$1661, 1666. As 
mentioned in \S \ref{sec:sample}, while the EW values of Ly$\alpha$ are similar between the non-AGN and the 
AGN composite spectra, the strongest emission line indicative of an AGN in the non-AGN spectra is 
\ion{C}{4} $\lambda$1549, which is only seen at 1\% the strength of Ly$\alpha$.

The ratios of strong UV emission lines have been used to understand the underlying shock and
photoionization conditions in the gas. Ly$\alpha$ is generally not used for diagnostic purposes because of the 
resonance scattering and the effects of strong dust attenuation on this line. The ratios of 
\ion{C}{4} $\lambda$1549/\ion{C}{3}] $\lambda$1909, \ion{C}{4} $\lambda$1549/\ion{He}{2} $\lambda$1640,
and \ion{C}{3}] $\lambda$1909/\ion{He}{2} $\lambda$1640 can be used to discriminate between
shock and photoionization predictions for narrow emission-line regions in AGNs \citep{villar1997, 
allen1998, groves2004}.  For the full composite spectrum we measure values of the ratios of
these lines that are very similar to the range of observed ratios in high-redshift radio galaxies 
\citep{villar1997,nagao2006b,matsuoka2009}.

\subsection{Absorption Features}
\label{sec:absfeat}

In addition to the strong emission lines seen in the composite spectrum, we detect several 
weak interstellar absorption features as well, such as the low-ionization lines \ion{Si}{2} $\lambda$1260,
\ion{O}{1} $\lambda$1302+\ion{Si}{2} $\lambda$1304, and \ion{Si}{2} $\lambda$1527. We do not
detect \ion{C}{2} $\lambda$1334 in our composite spectrum, which is a low-ionization absorption feature
detected in the non-AGN composite spectrum from \citet{shapley2003}\footnote{While not detected in the
full composite spectrum, \ion{C}{2} $\lambda$1334 is marginally detected in four of the 
individual spectra, HDF-BX160, Q0000-C14, Q0100-BX172, and Q1700-MD157.}.
The origin of this difference requires further study. The unblended low-ionization lines detected do 
not show significant blueshifts, in contrast to the blueshifts of $\sim 150 - 200$ km s$^{-1}$ observed 
in the non-AGN sample \citep{shapley2003,steidel2004,steidel2010}. This conclusion is dependent on our 
definition of the rest frame for the individual spectra. As discussed
in \S \ref{sec:generating}, our method uses the centroid of \ion{He}{2} $\lambda$1640 emission as a proxy
for the rest frame. On the other hand, the rest frame for the spectra included in the non-AGN LBG 
composite from \citet{shapley2003} was established using average relationships between the centroids of
rest-frame optical nebular emission lines and those of rest-frame UV features, such as interstellar 
absorption lines and Ly$\alpha$ emission \citep{adelberger2003}. Below we discuss how a change in 
the rest-frame determination would affect our conclusions about interstellar absorption line 
kinematics. However, we note that, regardless of the rest-frame determination, the observed offset 
between Ly$\alpha$ emission and low-ionization interstellar absorption lines in the AGN composite 
spectrum ($\sim 200$ km s$^{-1}$) is significantly smaller than the corresponding offset observed in
the non-AGN composite ($\sim 510$ km s$^{-1}$) -- indicating an intrinsic difference in kinematics 
traced by low-ionization gas. While the observed centroid of \ion{O}{1} $\lambda$1302+\ion{Si}{2} 
$\lambda$1304 may indicate a blueshift, it is most likely contaminated by additional absorption at 
1296 \AA. This absorption may be due to the stellar feature \ion{Si}{3} $\lambda$1296 
\citep{chandar2005}, but higher signal-to-noise spectra will be required to confirm this possibility. 

	\begin{figure}
	\epsscale{1.2} 
	\plotone{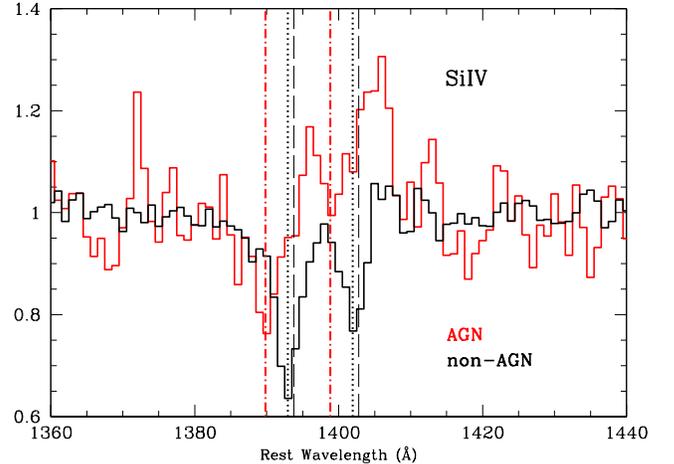} 
	\caption{Comparison of the blueshift of the \ion{Si}{4} $\lambda$1393 and $\lambda$1402 line 
	between the continuum normalized AGN composite spectrum (in red) and the non-AGN composite 
	spectrum (in black) presented in \citet{shapley2003}. The long dashed lines mark the rest-frame
	wavelengths of both \ion{Si}{4} $\lambda$1393 and $\lambda$1402, while the dotted and dot-dashed
	lines show the measured blueshifted centroids for the non-AGN and AGN composite
	spectra, respectively. The measured centroid of the 
	\ion{Si}{4} $\lambda$1393 feature indicates a velocity offset of $-$845$\pm$171 km s$^{-1}$ 
	in the AGN composite spectrum, while in the non-AGN spectrum, it is $-$180 km s$^{-1}$. Although
	it is difficult to determine robustly the full \ion{Si}{4} absorption profile in the AGN
	composite spectrum due to contamination by \ion{Si}{4} emission, we emphasize differences in the
	blue wing of the $\lambda$1393 absorption feature. There is excess absorption at the most
	blueshifted velocities in the AGN composite relative to that of the non-AGNs.
	\label{fig:SiIVcomp}} 
	\epsscale{1.2}
         \end{figure}

We also detect the high-ionization feature \ion{Si}{4} $\lambda \lambda$1393, 1402, both in 
emission and absorption. We use the stronger \ion{Si}{4} $\lambda$1393 feature as a probe of 
absorption line kinematics. This feature is resolved, with a FWHM greater than 500 km s$^{-1}$. 
The \ion{Si}{4} centroid indicates a significant blueshift, $-845 \pm 178$ km s$^{-1}$, which is 
several times larger than the one measured in the non-AGN composite spectrum ($\Delta v = -180$ km s$^{-1}$).
This difference is demonstrated in Figure~\ref{fig:SiIVcomp}, which shows both the AGN composite spectrum described 
here and the non-AGN composite from \citet{shapley2003}, zoomed in on the region around \ion{Si}{4}.
One potential cause for concern is the contamination of \ion{Si}{4} absorption profiles by the 
corresponding \ion{Si}{4} emission lines, given their close proximity in wavelength. The 
\ion{Si}{4} $\lambda$1393 emission line is in fact measured to be narrower than the $\lambda$1402
feature, suggesting that its blue edge is attenuated by \ion{Si}{4} $\lambda$1393 absorption, which 
is stronger than absorption from \ion{Si}{4} $\lambda$1402. Furthermore, the \ion{Si}{4} $\lambda$1393 
emission feature is observed to be weaker than that of \ion{Si}{4} $\lambda$1402, while the relative
oscillator strengths of the doublet members suggest that \ion{Si}{4} $\lambda$1393 should be twice 
as strong as $\lambda$1402. Both of these observations indicate the possible blending of \ion{Si}{4}
emission and absorption.

In order to quantify the contamination from emission in the observed 
\ion{Si}{4} $\lambda$1393 absorption profile, we require a robust model of the intrinsic 
\ion{Si}{4} $\lambda$1393 emission. Such a model is not straightforward to construct, given that 
both \ion{Si}{4} $\lambda$1393 and $\lambda$1402 emission features are affected by absorption, and, 
furthermore, the \ion{Si}{4} $\lambda$1402 emission feature is blended with emission from 
\ion{O}{4}] $\lambda$1401. The ratio between \ion{O}{4}] $\lambda$1401 and \ion{Si}{4} $\lambda$1402
is not well constrained, but can exceed unity \citep{hamann2002,nagao2006}, in which case the 
inferred \ion{Si}{4} $\lambda$1393 would be much weaker than if all the emission at $\sim 
1402$ \AA\ was due to \ion{Si}{4}. While the potential contamination from \ion{Si}{4} emission 
prevents us from tracing out the full \ion{Si}{4} absorption profile, we can still assert that absorbing
gas exists at a blueshift with a magnitude of $845$~km~s$^{-1}$. 
Furthermore, as shown in Figure~\ref{fig:SiIVcomp}, this highly blueshifted gas in the AGN composite spectrum at 
$\sim -1000$~km~s$^{-1}$ does not have a corresponding component in the non-AGN composite spectrum, 
representing a quantitative difference between the velocity profiles of outflowing absorbing gas 
in the two samples. Here we return to the issue of the discrepancy between the low-ionization 
\ion{Si}{2} absorption-line blueshift measured in the AGN and non-AGN LBG composite spectra. 
The non-AGN LBG composite spectrum has low-ionization \ion{Si}{2} absorption lines blueshifted by
$\sim 150 - 200$ km s$^{-1}$, while the \ion{Si}{2} absorption in the AGN composite spectrum does 
not show the same blueshift. If this difference were due to systematics of the rest-frame 
determinations, and we instead forced the \ion{Si}{2} absorption lines in the AGN composite to 
have the same blueshifts as those in the non-AGN composite, the blueshift of \ion{Si}{4} would 
\emph{increase} to $\geq 1000$ km s$^{-1}$.

\subsection{Continuum Shape} 
\label{sec:contshape}

	\begin{figure*}
	\epsscale{1.18} 
	\plotone{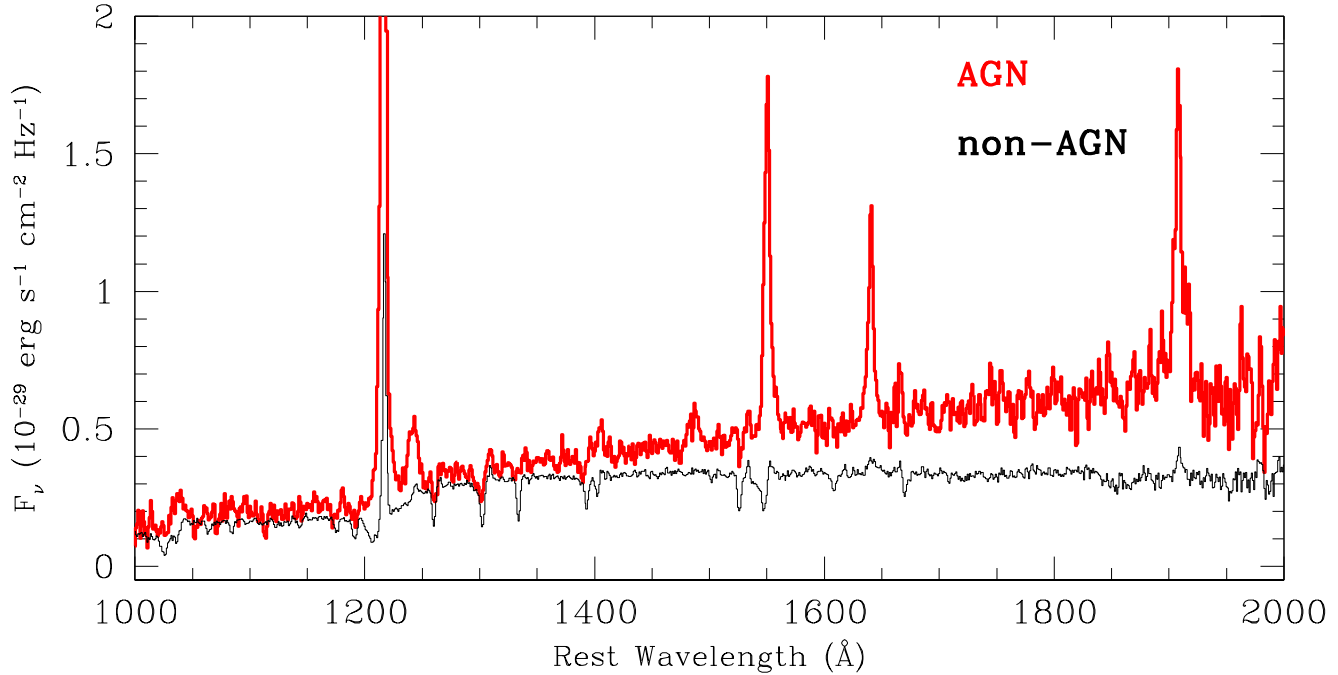} 
	\caption{Comparison of the continuum slope between the AGN composite spectrum and the non-AGN
	composite spectrum from \citet{shapley2003}. The AGN composite is significantly redder than
	the non-AGN composite. The power law slope $\beta$, as measured directly from the AGN spectrum, 
	is $\beta = -0.314$, while for the non-AGN spectrum, $\beta = -1.49$.  
	\label{fig:colorcomp}} 
	\epsscale{1.18}
         \end{figure*}

Besides the emission and absorption features seen in the composite spectrum, the shape of the
underlying continuum can be used to discern the presence of dust in these objects. The continuum
shape is is often described by $\beta$, the slope of a power law of the form $F_{\lambda}
\propto \lambda^{\beta}$ fit to the continuum. We used a modified version of the method outlined 
in \citet{calzetti1994}, using windows of the continuum with little to no activity as anchor
points for fitting the power law. For the full composite spectrum, we measured $\beta = -0.314$, 
while for the full non-AGN LBG composite from \citet{shapley2003}, we measured $\beta = -1.49$,
which is significantly bluer. A comparison between AGN and non-AGN composite spectra is shown 
in Figure \ref{fig:colorcomp}. We note here that in a composite spectrum of the control sample of 
objects at $2.17 \leq z \leq 2.48$, for which Ly$\alpha$ emission does not affect the $UG{\cal R}$ 
photometry, the UV slope is at least as red as in the composite spectrum for our full AGN sample. 
This result demonstrates that the red UV continuum slope measured for our AGN sample does not simply 
arise from red objects, which wouldn't satisfy the rest-frame UV selection criteria otherwise, 
scattering into the UV selection box due to the presence of strong Ly$\alpha$ emission. A full 
interpretation of the red AGN continuum awaits detailed population synthesis modeling which will be 
presented in Hainline et al. (in preparation), yet preliminary analysis suggests the AGN continuum 
does not contribute significantly at rest-frame UV wavelengths \citep{assef2009}. 

\citet{aird2010} argue that the redder rest-frame UV continuum slopes of UV-selected narrow lined 
AGNs relative to star-forming galaxies can result in a reduced overall selection efficiency in color-color
space. However, these authors do not take into account the simultaneous effects of continuum shape 
and strong emission lines in calculating the AGN selection efficiency, which can have a significant 
impact on the colors of narrow-lined AGNs in certain redshift ranges (as discussed in \S \ref{sec:sample}).
Our main conclusions about rest-frame UV spectroscopic trends within the UV-selected narrow-lined 
AGN sample are not significantly affected by these considerations. If anything, the difference we 
report between AGN and non-AGN continuum slope would be increased by including the spectra of red 
AGNs that do not fall in the color selection box due to their $G-{\cal R}$ colors.

\section{AGN Spectroscopic Trends}
\label{sec:trends}

In order to discern spectroscopic trends in the AGN sample, we split the objects into various
subsamples based on the properties of the individual spectra. Given the relatively small sample 
size (33 objects) and our desire to maximize the S/N of the resulting subsample composite spectra, 
we simply divide the sample in half for these analyses. The properties used to divide our sample
include: Ly$\alpha$ emission line EW, rest-frame UV absolute magnitude, and redshift. 
We do not split our sample by the EW for any of the other emission or absorption 
lines because most of our spectra do not have high enough S/N to allow for robust measurements of 
these features. The results for measuring various properties from the composite spectra 
described in this section are found in Table \ref{tab:binning}. 

\subsection{Ly$\alpha$ Dependencies}
\label{sec:lyadep}

Ly$\alpha$ is detected in emission in all of our AGN, with EW values that range from 
$W_{\mathrm{Ly\alpha}} = 10$\AA\ to $W_{\mathrm{Ly\alpha}} = 300$\AA. In contrast, the non-AGNs show Ly$\alpha$ 
both in emission and absorption. The median value for the Ly$\alpha$ EW in our sample was $W_{\mathrm{Ly\alpha}} = 63$\AA,
which is where we split the sample, creating a composite spectrum for those spectra with 
$W_{\mathrm{Ly\alpha}} < 63$\AA\ and those with $W_{\mathrm{Ly\alpha}} > 63$\AA. The results of measuring
the various properties of the composite spectra are found in Table \ref{tab:binning}.
The measured Ly$\alpha$ EW in the strong Ly$\alpha$ composite spectrum is $W_{\mathrm{Ly\alpha, strong}} = 123 \pm 14$\AA,
which is almost five times larger than what is measured in the weak Ly$\alpha$ composite spectrum, 
$W_{\mathrm{Ly\alpha, weak}} = 28 \pm 5$\AA. The Ly$\alpha$ EWs measured from the composite spectra
are consistent with the sample mean values of $\langle W_{\mathrm{Ly\alpha, strong}} \rangle = 141 \pm 15$\AA\
and $\langle W_{\mathrm{Ly\alpha, weak}} \rangle = 33 \pm 4$\AA. 

The strong Ly$\alpha$ composite spectrum has significantly
larger \ion{C}{4}, \ion{He}{2}, and \ion{C}{3}] EW values than the spectrum with weaker Ly$\alpha$
emission, but only by a factor of two. The same trend is observed when we consider line fluxes as
opposed to equivalent widths. For both equivalent widths and fluxes, the change in Ly$\alpha$ is not 
accompanied by an equivalent change in the strength of emission lines that are primarily sensitive 
to the level of nuclear activity.

\begin{figure*}
	\epsscale{1.18} 
	\plotone{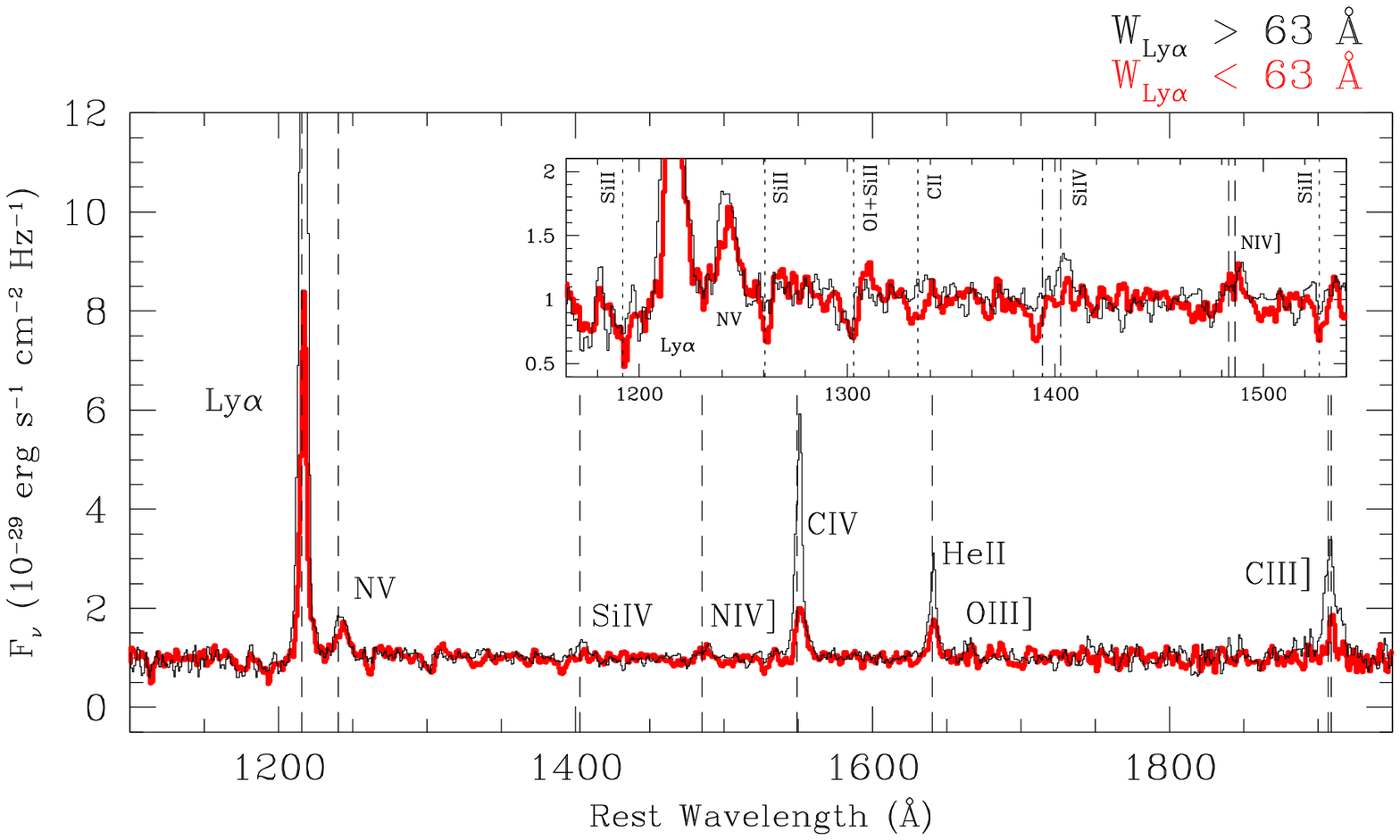} 
	\caption{Composite spectra of objects separated by Ly$\alpha$ EW. Plotted in black is 
	the continuum normalized composite spectrum from the half of galaxies in our AGN sample with the 
	strongest Ly$\alpha$ emission, while plotted in red is the continuum normalized composite 
	spectrum from the half of galaxies with the weakest Ly$\alpha$ emission. The inset shows a 
	section of the spectra between 1250 and 1540 \AA. The absorption lines 
	are significantly weaker in the strong Ly$\alpha$ composite spectrum, which is consistent 
	with the results from \citet{shapley2003} for non-AGN LBGs. Dashed lines indicate
	emission lines, dotted lines indicate low-ionization absorption features, and
	dot-dashed lines indicate high-ionization absorption lines. 
	\label{fig:LyAsplit}} 
	\epsscale{1.}
         \end{figure*}

In the strong Ly$\alpha$ EW composite spectrum, most low-ionization (\ion{Si}{2} $\lambda$1260,
\ion{C}{2} $\lambda$1334, and \ion{Si}{2} $\lambda$1527) and high-ionization (\ion{Si}{4} $\lambda 
\lambda$1393, 1402) absorption lines are statistically weaker than the features measured 
in the weak Ly$\alpha$ EW composite spectrum. This difference is shown in Figure \ref{fig:LyAsplit},
which overplots the strong and weak Ly$\alpha$ EW composite spectra. This trend of 
absorption line strength being anticorrelated with Ly$\alpha$ EW is also seen in the non-AGN LBGs \citep{shapley2003}. 
The \ion{O}{1} $\lambda$1302+\ion{Si}{2} $\lambda$1304 absorption line, however, is of a similar 
strength in both spectra. The potential contamination by \ion{Si}{3} $\lambda$1296 discussed in 
Section \ref{sec:absfeat} is more prominent in the strong Ly$\alpha$ EW composite, while it is not 
seen in the weak Ly$\alpha$ EW composite. In the weak Ly$\alpha$ EW composite, the \ion{O}{1}+\ion{Si}{2}
feature has a much smaller offset than what is seen in the full composite spectrum. Finally, \ion{C}{2}
is detected in the weak Ly$\alpha$ spectrum while not present in either the strong Ly$\alpha$ or full 
composite spectra (see \S \ref{sec:absfeat}). Beyond the trends in emission and absorption line 
strength, the two binned spectra do not differ in many other respects. They are not statistically 
different in UV luminosity or continuum color. 

\subsection{UV Magnitude Dependencies}

	\begin{figure}
	\epsscale{1.2} 
	\plotone{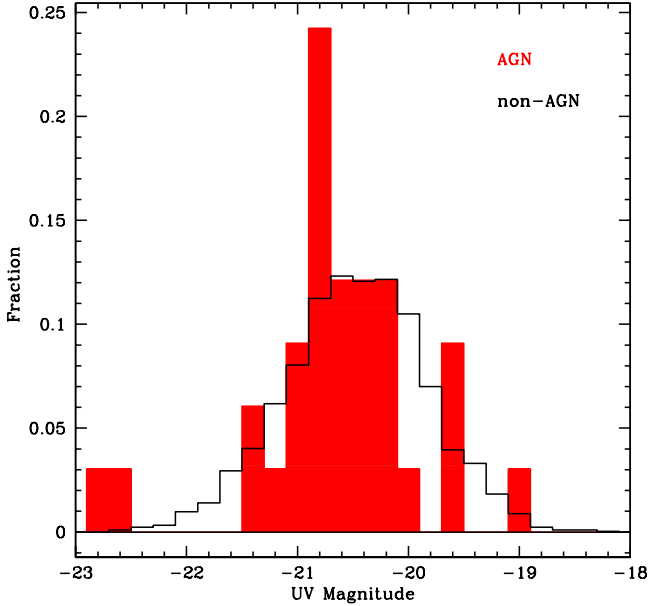} 
	\caption{The normalized UV absolute magnitude distribution for our sample of UV-selected 
	narrow-lined AGNs, compared to the distribution for the $z \sim 2-3$ UV-selected non-AGN sample.
	These UV magnitudes were calculated from the $G-{\cal R}$ colors and Ly$\alpha$-derived redshifts 
	for the objects.
	\label{fig:UVMag}} 
	\epsscale{1.2}
         \end{figure}

We calculated the UV magnitude for each of our galaxies using the value for a galaxy's redshift 
to interpolate the flux at 1500 \AA\ from the $G-R$ color. With this flux, we calculated a UV
luminosity and AB magnitude for each object. The normalized UV magnitude distributions for the AGN 
and non-AGN samples are shown in Figure \ref{fig:UVMag}. The median magnitude is $M_{UV,AB} = -20.7$, so
we split the sample into $M_{UV,bright}$ and $M_{UV,faint}$ subsamples for making composite spectra. 

The resulting spectra have Ly$\alpha$ EWs that are statistically similar, while the $M_{UV,faint}$
sample has overall weaker interstellar absorption lines. This difference in interstellar absorption
line strength is not as significant as the one observed between the two Ly$\alpha$ composite spectra
described in \S \ref{sec:lyadep}, indicating a stronger connection between Ly$\alpha$ emission strength
and interstellar absorption strength, as observed in non-AGN LBGs \citep{shapley2003}. Additionally, 
the two spectra separated by $M_{UV}$ are quite 
different in continuum color. The M$_{UV,bright}$ composite spectrum is significantly bluer 
($\beta_{spec} = -0.67$) than the M$_{UV,faint}$ composite ($\beta_{spec} = -0.06$), 
yet both are redder than non-AGN LBG composite ($\beta_{spec} = -1.49$). The average line 
flux for AGN-sensitive features such as \ion{C}{4}, \ion{He}{2}, and \ion{C}{3}] is larger 
in the M$_{UV,bright}$ composite than in the M$_{UV,faint}$ one. If the line fluxes 
and UV continuum flux density scaled by the same factor, which would indicate a direct connection 
between nuclear activity and UV continuum emission, the average EWs in the bright and faint 
composites would also be identical. In fact, the average EWs of the AGN-sensitive features in the M$_{UV,faint}$
sample are larger than those in the M$_{UV,bright}$ sample, suggesting that the relationship between
AGN-sensitive emission lines and UV continuum is not direct. While this trend is consistent with the
trend observed in broad-lined QSOs \citep{baldwin1977}, as these are narrow-lined 
AGNs, with only an obscured view of the central engine, we do not expect that the AGN emission contributes to
the overall UV continuum \citep{assef2009} in either the bright or faint subsample. The difference in
emission-line strength may simply indicate the correlation of AGN activity with additional properties of 
the host galaxy stellar population. Further population synthesis modeling is required to fully characterize 
these trends.

\subsection{Redshift Dependencies}

We separated the spectra by redshift at $z = 2.7$, effectively splitting between the $z \sim 3$ LBG
sample and the newly added $z \sim 2$ BX/BM and MD objects. There are 9 objects used to 
generate the $z > 2.7$ composite, and 24 to generate the $z < 2.7$ composite. The results for 
separating the spectra by redshift are very similar to those seen when separating
by $W_{Ly\alpha}$, with the higher redshift sample showing stronger Ly$\alpha$ emission on average. 
The Ly$\alpha$ EW distribution for the high redshift sample spans the range of 15 to 300 \AA\ 
($\langle W_{Ly\alpha, z > 2.7} \rangle = 124 \pm 28$), while the Ly$\alpha$ EW distribution for 
the low redshift sample spans the range of 10 to 145 \AA\ ($\langle W_{Ly\alpha, z < 2.7} \rangle = 
62 \pm 8$). Given that strong line emission can affect the selection of these objects as a
function of redshift, we removed from the analysis the 5 AGNs whose broadband colors only satisfied
the $UG{\cal R}$ selection criteria because of the presence of Ly$\alpha$ or \ion{C}{4}. Even with
these objects removed we still find a larger average Ly$\alpha$ EW for the higher redshift sample.
While these differences are suggestive of redshift evolution in the Ly$\alpha$ EW distributions
in a similar sense to what is observed in star-forming galaxies over the same redshift range 
\citep{reddy2008,nilsson2009}, the small sample size precludes us from drawing any firm conclusions.

\section{Discussion}
\label{sec:discussion}

\begin{deluxetable*}{lrrrrrr}
\tabletypesize{\scriptsize}
\tablecaption{Spectroscopic Properties of Composite Spectrum Subsamples\label{tab:binning}}
\tablewidth{0pt}
\tablehead{
  & \colhead{$W_{Ly\alpha} > 63$} & \colhead{$W_{Ly\alpha} < 63$} & \colhead{$M_{UV,bright}$} & \colhead{$M_{UV,faint}$} & \colhead{$z > 2.7$} & \colhead{$z < 2.7$}
}
\startdata

$\mathrm{N_{gal}}$ & 16\phm{,0000} & 17\phm{,0000} & 16\phm{,0000} & 17\phm{,0000} & 9\phm{,00000} & 24\phm{,0000} \\
$\mathrm{\langle z_{Ly \alpha}\rangle \tablenotemark{a}}$                   & 2.60$\pm$0.08\phm{0}         & 2.49$\pm$0.06\phm{0}    & 2.65$\pm$0.08\phm{0}    & 2.43$\pm$0.05\phm{0}    & 2.94$\pm$0.07\phm{0}    & 2.37$\pm$0.02\phm{0} \\
$\mathrm{\langle M_{UV} \rangle \tablenotemark{a}}$                         & $-$20.6$\pm$0.2\phm{00} & $-$20.7$\pm$0.2\phm{00}    & $-$21.2$\pm$0.1\phm{00}    & $-$20.2$\pm$0.1\phm{00}    & $-$21.1$\pm$0.2\phm{00}    & $-$20.5$\pm$0.1\phm{00} \\
$\mathrm{\langle} W_{\mathrm{Ly\alpha}} \mathrm{\rangle \tablenotemark{a}}$ & 141$\pm$15\phm{.00}     & 33$\pm$4\phm{.000}     & 78$\pm$14\phm{.00}    & 82$\pm$17\phm{.00}    & 124$\pm$28\phm{.00} & 62$\pm$8\phm{.000} \\
$\mathrm{\langle \beta_{G-{\cal R}} \rangle \tablenotemark{a}}$             & $-$0.8$\pm$0.2\phm{00}  & $-$1.2$\pm$0.1\phm{00} & $-$1.1$\pm$0.2\phm{00} & $-$0.9$\pm$0.2\phm{00} & $-$1.1$\pm$0.4\phm{00} & $-$1.0$\pm$0.1\phm{00} \\
$\mathrm{\langle \beta_{spec} \rangle \tablenotemark{a}}$                   & $-$0.5$\pm$0.3\phm{00}  & $-$0.6$\pm$0.3\phm{00} & $-$0.9$\pm$0.3\phm{00} & $-$0.3$\pm$0.3\phm{00} & $-$0.9$\pm$0.4\phm{00} & $-$0.4$\pm$0.2\phm{00} \\
$\mathrm{\beta_{spec}\tablenotemark{b}}$                                    & $-$0.1$\pm$0.4\phm{00}   & $-$0.5$\pm$0.3\phm{00} & $-$0.7$\pm$0.3\phm{00} & $-$0.1$\pm$0.4\phm{00} & $-$0.7$\pm$0.5\phm{00} & $-$0.2$\pm$0.3\phm{00} \\
\\
$W_{\mathrm{Ly\alpha}}\tablenotemark{c}$     & 123$\pm$14\phm{.00}       & 28$\pm$5\phm{.000}      & 62$\pm$18\phm{.00}       & 56$\pm$16\phm{.00}     & 103$\pm$37\phm{.00}     & 52$\pm$8\phm{.000} \\
$W_{\mathrm{NV,1240}}\tablenotemark{c}$      & 6.23$\pm$1.54\phm{0}      & 4.42$\pm$1.01\phm{0}    & 3.22$\pm$0.84\phm{0}     & 8.21$\pm$1.80\phm{0}   & 4.15$\pm$1.66\phm{0}    & 5.83$\pm$1.12\phm{0} \\
$W_{\mathrm{NIV],1484}}\tablenotemark{c}$    & 2.13$\pm$0.86\phm{0}      & 1.68$\pm$0.53\phm{0}    & 1.05$\pm$0.52\phm{0}     & 2.51$\pm$1.17\phm{0}   & 3.07$\pm$1.36\phm{0}    & 1.57$\pm$0.58\phm{0} \\
$W_{\mathrm{CIV,1549}}\tablenotemark{c}$     & 25.21$\pm$4.15\phm{0}     & 7.99$\pm$2.56\phm{0}    & 10.17$\pm$3.84\phm{0}    & 16.40$\pm$4.15\phm{0}  & 40.43$\pm$8.06\phm{0}   & 12.01$\pm$2.89\phm{0} \\
$W_{\mathrm{HeII,1640}}\tablenotemark{c}$    & 9.16$\pm$1.88\phm{0}      & 5.48$\pm$1.71\phm{0}    & 6.01$\pm$1.23\phm{0}     & 8.13$\pm$3.09\phm{0}   & 14.03$\pm$5.29\phm{0}   & 6.27$\pm$1.41\phm{0} \\
$W_{\mathrm{CIII],1909}}\tablenotemark{c}$   & 20.85$\pm$8.00\phm{0}     & 6.06$\pm$2.66\phm{0}    & 14.08$\pm$5.90\phm{0}    & 15.62$\pm$6.74\phm{0}  & 36.75$\pm$15.58         & 8.75$\pm$2.90\phm{0} \\
\\
$W_{\mathrm{SiII,1260}}\tablenotemark{c}$    & -\phm{00000}              & $-$1.81$\pm$0.49\phm{0} & $-$0.87$\pm$0.34\phm{0}  & -\phm{00000}           & -\phm{00000}            & $-$1.80$\pm$0.39\phm{0} \\
$W_{\mathrm{OI+SiII,1303}}\tablenotemark{c}$ & $-$2.00$\pm$0.82\phm{0}   & $-$1.96$\pm$0.54\phm{0} & $-$2.36$\pm$0.47\phm{0}  & -\phm{00000}           & -\phm{00000}            & $-$2.27$\pm$0.54\phm{0} \\
$W_{\mathrm{SiIV,1393}}\tablenotemark{c}$    & -\phm{00000}              & $-$1.72$\pm$0.74\phm{0} & $-$1.66$\pm$0.55\phm{0}  & -\phm{00000}           & -\phm{00000}            & $-$1.34$\pm$0.67\phm{0} \\
$W_{\mathrm{SiII,1527}}\tablenotemark{c}$    & -\phm{00000}              & $-$1.03$\pm$0.53\phm{0} & $-$0.85$\pm$0.35\phm{0}  & -\phm{00000}           & -\phm{00000}            & $-$0.79$\pm$0.35\phm{0} \\

\enddata
\tablenotetext{a}{Sample average values for the composite spectra.}
\tablenotetext{b}{UV-continuum slope, measured from the composite spectra.}
\tablenotetext{c}{Rest-frame EW in \AA, measured from the composite spectra. Positive values indicate 
emission, while negative values indicate absorption. Uncertainties are calculated as described in
\S \ref{sec:obsdata}. This table includes absorption and emission measurements with greater than
2$\sigma$ significance.}

\end{deluxetable*}

Analysis of the composite spectrum of the UV-selected AGNs at $z \sim 2 - 3$ reveals a number
of results regarding the nature of AGN activity. We report the detection of weak absorption lines from both low- and
high-ionization species. Most strikingly, the high-ionization \ion{Si}{4} absorption feature 
exhibits a significant blueshift of $\Delta v = -845\pm178$ km s$^{-1}$. The precise value of this
blueshift is referenced to our estimate of the rest frame based on \ion{He}{2} $\lambda$1640. As
discussed in \S \ref{sec:absfeat}, if we adopt a rest frame in which the low-ionization lines have 
the same blueshift as those in the non-AGN LBG composite spectrum of \citet{shapley2003}, then the
magnitude of the inferred blueshift for \ion{Si}{4} would be even greater. While contamination from \ion{Si}{4} emission 
prevents us from tracing out the full \ion{Si}{4} velocity profile in absorption, the most strongly 
blueshifted material appears to be outflowing more rapidly than the associated gas in star-forming 
galaxies at $z \sim 2 - 3$ \citep{shapley2003,steidel2010}. \citet{thacker2006} presents results of 
a simulation that indicate that pure star formation, even when the full kinetic energy from each 
supernovae is applied to the outflowing material, can only produce maximum outflow velocities of 
roughly $v = 600$ km s$^{-1}$. Their modeling shows that only AGNs and quasars can produce high 
speed outflows with velocities greater than $10^{3}$ km s$^{-1}$. Outflows of this magnitude have 
also been seen in poststarburst galaxies at $z = 0.6$, using \ion{Mg}{2} $\lambda \lambda$2796, 2803, 
which are claimed to result from the effects of an AGN \citep{tremonti2007}. 

Previous studies of outflows observed in galaxies with AGNs can be used to place these 
UV-selected AGN results in context. \citet{krug2010} presented a study of outflows from a sample 
of local IR-faint AGN. For the narrow-lined objects, the outflow velocities (as calculated from 
the \ion{Na}{1} D interstellar absorption line doublet) are on the order of those from starburst 
galaxies, offering a conclusion that  star formation was the process driving the outflows in 
Seyfert 2 systems. Based on a sample of local infrared-luminous starburst galaxies exhibiting AGN
activity, \citet{rupke2005} show evidence for high velocity superwinds, which they compare to those from a 
non-AGN ULIRG sample presented in \citet{rupke2005b,rupke2005c}. Both have comparable outflow 
velocities, leading to the conclusion that the momentum and energy required for the outflow could 
have come equally from a starburst or the AGN. The current work represents the same type of 
differential comparison between AGNs and their non-active counterparts, but at high redshift, 
when the black hole and bulge are both actively forming. This analysis highlights the specific
effects of the AGN on the outflowing ISM. Relative to work on AGN outflows at high redshift, 
which has focused on the extended line emission in individual systems or small samples of AGNs 
alone \citep{alexander2010, nesvadba2008}, the benefit of our analysis lies in our comparison to 
a control sample of non-AGN star-forming galaxies.

In order to understand how the outflows observed in our sample of AGNs will ultimately affect the 
galactic gas content, a calculation of the mass outflow rate of the gas is needed. Such a calculation
requires knowledge of the outflowing gas metallicity, column density, covering fraction and physical location with 
respect to the illuminating source. Specifically, it is necessary to determine whether the outflowing
gas extends over the scale of the entire galaxy or is confined to the scale of the central engine.
Based on current data, we cannot obtain a precise estimate of the location of the gas, metallicity,
column density, or covering fraction. Without determinations of these properties, a full comparison 
to AGN feedback models \citep[e.g.][]{thacker2006} cannot be made. 

The \ion{N}{4}] $\lambda$1486 emission line is detected in our AGN composite. This feature is not
detected in the non-AGN LBG composite \citep{shapley2003}, and is observed only rarely in broad-lined
quasars in the Sloan Digital Sky Survey \citep{bentz2004,jiang2008}. At the same time, \ion{N}{4}] 
has been detected in the spectra of high redshift radio galaxies \citep{vernet2001,humphrey2008}. In
order to determine the origin of the \ion{N}{4}] emission, \citet{humphrey2008} consider both
photoionization and shock models. A comparison of model predictions for line ratios such as \ion{N}{4}] / 
\ion{C}{4} and \ion{N}{4}] / \ion{He}{2} with those measured in our composite spectrum
suggests that the observed \ion{N}{4}]  originates in photoionized gas. Our observed line ratios 
indicate that the gas is of solar or supersolar metallicity and subjected to a hard ionizing spectrum 
($f_\nu \propto \nu^{\alpha}$, where $\alpha \geq -1.0$) with ionization parameter $U \geq 0.05$, 
where $U$ is defined here as the ratio of ionizing photons to H atoms at the surface of the model 
photoionized gas slab. On the other hand, shock excitation models cannot explain our high observed 
values of \ion{N}{4}] / \ion{C}{4}. 

The EW of Ly$\alpha$ in the UV spectra of our AGNs is indicative of both the strength of the 
AGN as well as the properties of star-forming regions. The observed Ly$\alpha$ EW is further modulated 
by various radiative transfer effects, due to its high scattering cross section. In non-AGN LBGs,
the strength of Ly$\alpha$ emission has been shown to correlate with the EW of low-ionization 
interstellar absorption lines, such that strong Ly$\alpha$ emission is accompanied by weaker interstellar
absorption \citep{shapley2003}. This result can be understood if the escape of Ly$\alpha$ photons is at 
least partially modified by the covering fraction of neutral gas in the ISM. Additionally,
\citet{kornei2010} show that stronger Ly$\alpha$ emission is coupled with smaller dust obscuration as
traced by the slope of the UV continuum, indicating that interaction with dust preferentially destroys 
Ly$\alpha$ photons, a result consistent with previous work by \citet{shapley2003}, 
\citet{pentericci2007}, and \citet{verhamme2008}.

When separating our objects according to Ly$\alpha$ EW to create composite spectra, 
we find that the composite spectrum created from objects with large Ly$\alpha$ EW shows 
stronger \ion{C}{3}], \ion{C}{4} and \ion{He}{2} emission than the composite spectrum created from 
objects with smaller Ly$\alpha$ EWs. As shown in Table \ref{tab:binning}, the \ion{C}{3}], \ion{C}{4}, 
and \ion{He}{2} lines are two to three times weaker in the $W_{Ly\alpha} < 63$ composite spectrum than 
in the $W_{Ly\alpha} > 63$ composite spectrum. This result suggests that the strength of Ly$\alpha$
emission is modulated at least partially by the level of AGN activity, which is traced by the 
strength of these other emission lines. However, the fact that the Ly$\alpha$ EW is almost 5 times weaker in 
the weak Ly$\alpha$ EW composite spectrum indicates additional suppresion of Ly$\alpha$ photons
beyond the reduced level of AGN activity. At the same time, the low-ionization interstellar 
absorption lines that indicate the covering fraction of cool gas are significantly stronger in 
the weak Ly$\alpha$ composite. Therefore, it is not only the strength of the underlying AGN that 
separates the objects by Ly$\alpha$ EW, but also the covering fraction of gas that might absorb 
and reradiate the Ly$\alpha$ emission. This trend agrees with the results of \citet{shapley2003} 
for non-AGN LBGs.

In our AGN sample, the UV continuum of the strong Ly$\alpha$ composite is redder 
than that of the weak Ly$\alpha$ composite, though the difference is not significant. This discrepancy 
with the trends among non-AGN LBGs from \citet{shapley2003} and \citet{kornei2010} may be a result of 
our small sample size, or potentially because the Ly$\alpha$ flux we observe originates from
both the general star-forming ISM as well as the nuclear region. These two sources of Ly$\alpha$
photons may have disjoint properties with respect to the geometry of dust extinction, suppressing
the trend observed among the non-AGN LBGs. 

Future modeling of the spectral energy distributions of the AGN host galaxies (Hainline et al. in prep.)  
will allow for analysis of the UV spectra separated by stellar mass and E(B-V), as well 
as uncover spectral trends as a function of galaxy evolutionary state. We also will probe the origin 
of the strikingly red UV continuum slopes found in the narrow-line UV-selected AGN spectra.  

\acknowledgments 

We would like to thank Dawn Erb and Max Pettini for their helpful discussions. We would also like to
acknowledge the referee, David Alexander, for a thorough and constructive report, which significantly 
improved the paper. A.E.S. acknowledges support from the David and Lucile Packard Foundation. C.C.S. 
acknowledges additional support from the John D. and Catherine T. MacArthur Foundation and the Peter
and Patricia Gruber Foundation. We wish to extend special thanks to those of Hawaiian ancestry on 
whose sacred mountain we are privileged to be guests. Without their generous hospitality, most of the 
observations presented herein would not have been possible.

\bibliographystyle{apj}
\bibliography{apj-jour,lbgrefs}



\end{document}